\documentclass[11pt,twoside]{article}

\usepackage{asp2006}
\usepackage{epsf}
\usepackage{lscape}
\usepackage{graphicx}

\begin{document}
\title{AMIGA project: Active galaxies in a complete sample of isolated galaxies.}   %%% Fill in title
\author{J.~Sabater$^1$,  S. Leon$^{1,2}$, L. Verdes-Montenegro$^1$, U. Lisenfeld$^2$, J. Sulentic$^{1,4}$, S. Verley$^{1,3,5}$,}   %%% Fill in author names
\affil{$^1$Instituto de Astrof\'{\i}sica de Andaluc\'{\i}a -- CSIC (jsm@iaa.es), $^2$Instituto de RadioAstronom\'{\i}a Milim\'etrica (IRAM), $^3$Universidad de Granada, $^4$University of Alabama, $^5$LERMA-OBSPM,}    %%% Fill in author affiliations

\begin{abstract} %%% Abstract to run on from here.
The project AMIGA (Analysis of the interstellar Medium of Isolated GAlaxies)
provides a statistically significant sample of the most isolated galaxies in the
northern sky. Such a control sample is necessary to understand the role of the
environment in evolution and galaxy properties like the interstellar medium
(ISM), star formation and nuclear activity. The data is publicly released under
a VO interface at http://amiga.iaa.es/. One of our main goals is the study of
nuclear activity in non-interacting galaxies using different methods. We focus
on the well known radiocontinuum-far infrared (FIR) correlation in order to find
radio-excess galaxies which are candidates to host an active galactic nucleus
(AGN) and FIR colours to find obscured AGN candidates. We looked for the
existing information on nuclear activity in the V\'eron-Cetty catalogue and in the
NASA Extragalactic Database (NED). We also used the nuclear spectra from the 
Sloan Digital Sky Survey which allow us to determine the possible presence of 
an AGN and to study the properties of the underlying stellar populations. We
produced a final catalogue of AGN-candidate galaxies which will provide a baseline 
for the study of the nuclear activity depending on the environment. We find that 
the fraction of FIR selected AGN-candidates ranges between 7\% and 20\%. There are 
no radio-excess galaxies in our sample above a factor 5 of radio excess which is 
the lowest rate found in comparison with other samples in denser environments. 
Finally, we obtained a fraction of about 22\% of AGN using the optical spectra, 
a significant fraction for a sample of isolated galaxies. We conclude that the 
environment plays a crucial and direct role in triggering radio nuclear activity 
and not only via the density-morphology or the density-luminosity relations.
\end{abstract}

\section{Introduction}

Galaxy evolution depends strongly on the environment. In particular, galaxy-galaxy interactions can induce nuclear activity by removing angular momentum from the gas and, in this way, feeding the central black hole. Hence, a higher rate of nuclear activity would be expected in interacting galaxies. However, different studies of this topic lead to contradictory results.
In order to understand the role of the environment in the nuclear activity we need a statistically significant sample of isolated galaxies which will act as the baseline for this study. The study was developed in the frame of the AMIGA project (for a detailed description of AMIGA see the contribution of J. Sulentic in this proceedings). 

\section{Selection of active galaxies}

We applied a completeness test known as $<V/V_m>$ as explained in \citet{V2005} and \citet{L2007} to the AMIGA isolated galaxies obtaining a complete subsample which contains 710 galaxies. This sample constitute the base for our statistical studies.

We made a catalogue of AGN candidates of isolated galaxies looking for data in the literature and using three selection methods: FIR colour, radiocontinuum-FIR correlation radio-excess and based on the optical spectra \citep{S2008,S2009}.

We did a cross-correlation of our sample with: a) The NED (NASA Extragalactic Database) database: 77 galaxies found, 22 of them AGN and b) V\'eron-Cetty \& V\'eron active galaxies catalogue (12th edition): 25 galaxies found, 18 of them AGN.

In the work of \citet{d1985} is shown a method to identify AGN candidates using FIR properties.
%A method to identify AGN candidates is to observe its FIR emission. 
Galaxies hosting an AGN have, in general, a flatter spectrum in FIR. 
%This is due the hotter temperatures of the dust warmed by the central engine. 
The advantage of this method is that it can find obscured AGN that cannot 
be observed using other wavelengths or methods. 
The success rate of the method is about 70\%. We select the galaxies with an spectral index between $25 \mu m$ and $60 \mu m$ of $\alpha_{25,60} > -1.958$. There are 58 AGN candidates which amounts to 7\% of the sample or 20\% of the galaxies classified with this method.

The correlation between the FIR and the radio continuum emission
is very tight and is attributed to star formation.
 \citep{C1991}. 
Deviation from this correlation may be produced by a radio-loud active nucleus. 
If the additional emission from the active nucleus is strong enough, an excess of radio
emission will be found with respect to the radio-FIR correlation.
%We use survival analysis methods to compute the correlation, obtaining $ \log L_{1.4 GHz}(W~Hz^{-1})= [1.03 \pm 0.03] \log (L_{FIR}/L_{\odot}) + [11.4 \pm 0.3]$. 
Radio-excess galaxies are the ones whose radio luminosity is larger than 5 times the value predicted by the radio-FIR correlation \citep{Y2001}. There are 7 radio-excess galaxies in the complete sample which amount $\sim 1\%$ of the sample. This is a very low rate.
There is a chance that the radio excess found
using the NVSS data is in fact due to a background/foreground source projected
in the line of sight of the galaxy. We estimated that 14 of the radio detections in our sample could be due to unrelated sources. In order to determine which sources were genuine detections and which not we obtained high resolution VLA radio continuum images of the radio-excess galaxies. Finally, we found that all the radio emission excess in the radio-excess sources was produced by unrelated sources. This leads to a final rate of radio-excess galaxies of 0\%.

%\begin{figure}
%\centering
%   \includegraphics[width=9.8cm]{lfir-lradio-total.eps}
%   \caption{Radio/FIR luminosity diagram for the total sample. We show the correlation as a continuum line and the 5 times radio-excess and FIR-excess level as dashed lines. The galaxies above the upper dashed line are the radio-excess galaxies.
%           }
%      \label{l}
%\end{figure}

We also used the nuclear spectra from the Sloan Digital Sky Survey. There are 353 spectra of AMIGA galaxies in the 6th Data Release. We obtained the underlying stellar populations using the Starlight code. Subtracting the estimated emission from the stellar populations we obtained the nebular emission and determined the presence of an AGN, star formation or a transition object (TO; with properties in between the former two types) using the typical diagnostic diagrams. We also determined the type of AGN when possible. Only $\sim7\%$ lack of emission lines, $\sim56\%$ are classified as star forming, $\sim16\%$ as TOs and $\sim22\%$ as AGN. This is a significant fraction for a sample of isolated galaxies. 

\section{Comparison with denser environments}

\begin{figure}
\centering
%   \resizebox{8.5cm}{8.5cm}{\includegraphics{alfa-selection.eps}}
   \includegraphics[width=12cm]{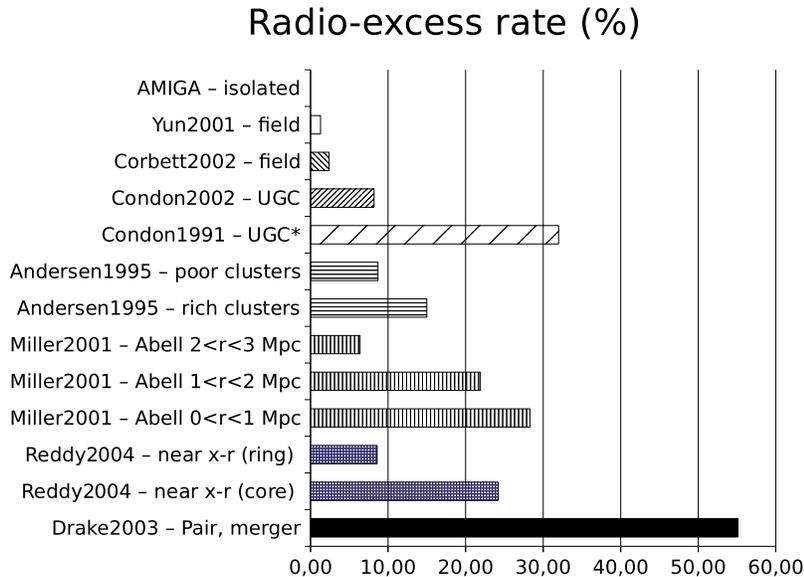}
   \caption{Percentage of galaxies above a factor 5 of radio excess in the comparison samples. Notice that the percentage for the AMIGA sample is 0\%.
           }
      \label{re}
\end{figure}

We compared the radio-excess rate with other samples in denser environments: those referred to as field galaxies in the literature
\citep[e.g.,][]{C1991b,Y2001,Co2002,C2002,D2003}
where usually no environmental selection criterion has been applied, and cluster
samples \citep[e.g.,][]{A1995,M2001,R2004}. The results are summarized in Figure~\ref{re}. We also took into account the possible effect of the luminosity as well a the morphology on the rate of radio-excess galaxies to avoid the biases coming from the morphology-density and luminosity-density relations. We found the same increase of the fraction of radio-excess galaxies toward denser environments independently of the luminosity or the morphology.

We compared the rates of nuclear activity derived from the optical spectra with the ones found by \citet{M2008} for two samples of galaxies in compact groups (CG). The reduction of the data and the measurement method were very similar. The results are summarized in Table~\ref{isol_cg}. We will take into account the morphology and luminosity of galaxies in a future to avoid possible biases.

\begin{table}
\caption[Fraction of AGN for isolated and compact groups]{Fraction of the different types of AMIGA galaxies and compact groups galaxies from \citet{M2008}. Fractions given in percentages with respect to the total number of galaxies.}
\begin{center}
\label{isol_cg}
\begin{tabular}{l r r r}
\hline
Sample & AMIGA & Hickson CG & UZC-CG\\
\hline \hline
Non-emission:             &   7.1 &  37.5 &  31.0 \\
Emission:             &  92.9 &  62.5 &  69.0 \\
~~SF                                     &  55.8 &  20.1 &  20.9 \\
~~TO                                      &  15.5 &  13.4 &  12.6 \\
~~AGN                                    &  21.7 &  29.0 &  35.5 \\
%~~~~NLAGN total:                          &  19.5 &  28.3 &  33.5 \\
%~~~~~~LINER                               &   8.9 &  13.0 &   2.8 \\
%~~~~~~Sy2                                 &   4.4 &   8.9 &  10.8 \\
%~~~~~~unclassified NLAGN (LLAGN)          &   6.2 &   6.3 &  19.9 \\
%~~~~Sy1                                   &   2.2 &   0.7 &   2.0 \\
%~~TO + AGN:                               &  37.2 &  42.4 &  48.1 \\
\hline
\end{tabular}
%\begin{list}{}{}
%\item[$^{1}$] UZC-CG subsample with spectra from FAST and the SDSS.
%\item[$^{4}$] Classified galaxies in the AMIGA sample and galaxies with emission in the compact groups sample.
%\end{list}
\end{center}
\end{table}

\section{Conclusions}

%We have produced a catalogue of AGN candidates for the CIG sample, therefore we can study nuclear activity in a complete low density well defined sample of isolated galaxies. The activity rate derived from the radio-FIR correlation is very low (0\%) in comparison with denser environments what shows us how environment is fundamental for the nuclear activity.

We
produced a final catalogue of AGN-candidate galaxies which will provide a baseline 
for the study of the nuclear activity depending on the environment. We find that 
the fraction of FIR selected AGN-candidates ranges between 7\% and 20\%. There are 
no radio-excess galaxies in our sample above a factor 5 of radio excess which is 
the lowest rate found in comparison with other samples in denser environments, 
independently of the luminosity
or the morphology. 
Finally, we obtained a fraction of about 22\% of AGN using the optical spectra, 
a significant fraction for a sample of isolated galaxies. 

We conclude that the 
environment plays a crucial and direct role in triggering radio nuclear activity 
and not only via the density-morphology or the density-luminosity relations.

%%% MAIN BODY OF TEXT GOES HERE. CONSULT "INSTRUCTIONS FOR AUTHORS USING
%%% LATEX2E MARKUP", SECTIONS 2.3-2.6 FOR HELP WITH EQUATIONS, FIGURES,
%%% AND TABLES.

%\section{}   %%% Top level section head (remove "%" symbol)
%\subsection{}   %%% Second level section head (remove "%" symbol)
%\subsubsection{}   %%% Lowest level section head (remove "%" symbol)
%\section*{}    %%% Unnumbered top level section head (remove "%" symbol)
%\subsection*{}   %%% Unnumbered second level section head (remove "%" symbol)

\acknowledgements %%% Text of acknowledgements runs on after this command.
We would like to warmly thank everybody who provided data for this study.
The authors are partially supported by DGI Grant
AYA2008-06181-C02 and Junta de Andaluc\'{\i}a (Spain) TIC-114 and P08-FQM-4205-PEX.

%%% THE BIBLIOGRAPHY
%%%
%%% CONSULT SECTION 3 OF "INSTRUCTIONS FOR AUTHORS" FOR HOW TO USE NATBIB.
%%% AUTHORS ARE ENCOURAGED TO USE EITHER THE "THEBIBLIOGRAPY" ENVIRONMENT
%%% BY UNCOMMENTING (DELETING THE "%" SYMBOL) THE COMMANDS BELOW, OR BY
%%% USING THE BIBTEX ENVIRONMENT. TO FIND OUT WHICH IS APPLICABLE TO YOUR
%%% CONTRIBUTION, CONSULT THE VOLUME EDITORS FOR YOUR PROCEEDINGS.
%%%


\begin{thebibliography}{}
\bibitem[Andersen \& Owen (1995)]{A1995} Andersen, V. \& Owen, F. 1995, AJ, 109, 1582
\bibitem[Condon et al. (1991)]{C1991} Condon, J., Anderson, M. \& Helou, G. 1991, ApJ, 376, 95
\bibitem[Condon \& Broderick (1991)]{C1991b} Condon, J. \& Broderick, J. 1991, AJ, 102, 1663
\bibitem[Condon et al. (2002)]{C2002} Condon, J., Cotton, W. \& Broderick, J. 2002, AJ, 124, 675
\bibitem[Corbett et al. (2002)]{Co2002} Corbett, E. et al. 2002, ApJ, 564, 650
\bibitem[Drake et al. (2003)]{D2003} Drake, C. et al. 2003, AJ, 126, 2237
\bibitem[de Grijp et al. (1985)]{d1985} de Grijp, M.~H.~K. et al. 1985, Nature, 314, 240
\bibitem[Karachentseva (1973)]{K1973} Karachentseva V. 1973, AIISAO, 8, 3
%\bibitem[Leon et al.(2008)]{L2008} Leon, S., et al.\ 2008, A\&A, 485, 475 
\bibitem[Lisenfeld et al.(2007)]{L2007} Lisenfeld, U., et al.\ 2007, A\&A, 462, 507
\bibitem[Martinez PhD (2008)]{M2008} Martinez, M.~A. 2008, PhD thesis, IAA-CSIC
\bibitem[Miller \& Owen (2001)]{M2001} Miller, N. \& Owen, F. 2001, AJ, 121, 1903
\bibitem[Reddy \& Yun (2004)]{R2004} Reddy, N. \& Yun, M. 2004, ApJ, 650, 695
\bibitem[Sabater et al. (2008)]{S2008} Sabater, J., et al.\ 2008, A\&A, 486, 73 
\bibitem[Sabater PhD (2009)]{S2009} Sabater, J. 2009, PhD thesis, IAA-CSIC
\bibitem[Sulentic et al. (2006)]{S2006} Sulentic et al. 2006, A\&A, 449, 937
\bibitem[Verdes-Montenegro et al. (2005)]{V2005} Verdes-Montenegro L. et al. 2005, A\&A, 436, 443
%\bibitem[Verley et al.(2007a)]{V2007a} Verley, S., et al.\ 2007a, A\&A, 470, 505 
%\bibitem[Verley et al.(2007b)]{V2007b} Verley, S., et al.\ 2007b, A\&A, 472, 121 
%\bibitem[V\'eron-Cetty \& V\'eron (2006)]{V2006} V\'eron-Cetty \& V\'eron 2006, A\&A, 455, 773
\bibitem[Yun et al. (2001)]{Y2001} Yun, M., Reddy, N., \& Condon, J. 2001, ApJ, 554, 803
%\bibitem[]{}


\end{thebibliography}
\end{document}